\documentclass[conference]{IEEEtran}
\IEEEoverridecommandlockouts

\usepackage{biblatex} 
\usepackage{fancyvrb}

\begin{document}

\title{STaKTAU: profiling HPC applications' \\
operating system usage}

\author{\IEEEauthorblockN{Camille Coti}
\IEEEauthorblockA{\textit{École de Technologie Supérieure}\\
Montreal, QC, Canada}
\and
\IEEEauthorblockN{Kevin Huck}
\IEEEauthorblockA{\textit{University of Oregon} \\
Eugene, OR, USA}
\and
\IEEEauthorblockN{Allen D. Malony}
\IEEEauthorblockA{\textit{University of Oregon} \\
Eugene, OR, USA}
}

\maketitle

\begin{abstract}
This paper presents a approach for measuring the time spent by HPC applications in the operating system's kernel. We use the SystemTap interface to insert timers before and after system calls, and take advantage of its stability to design a tool that can be used with multiple versions of the kernel. We evaluate its performance overhead, using an OS-intensive mini-benchmark and a raytracing mini app.
\end{abstract}

\section{Introduction}

Modern HPC applications are using a variety of programming interfaces and
environments, in complex, intertwined ways.
Optimizing application performance requires observing its execution behavior,
obtaining metrics its operation, and identifying opportunities for
performance improvement.
The metrics can include the time spent in code regions, data about
communication operation, hardware counters about cache use, and so on.

While most performance tools are focused on application-level observation,
the operating system (OS) plays an important role in modern
parallel execution environments.
For instance, how much time does the OS spend scheduling the
processes and threads, migrating memory between caches, allocating
memory, and doing IO operations
are key in understanding the whole performance picture?
Analyzing the time spent in system-level operations should be part
of the performance engineering process.

The problem is that this requires some means to observe the OS kernel's
operation.
A previous approach,
known as \textit{KTAU},
was integrated with the TAU toolset~\cite{Shende2006}
and was based on instrumenting
the Linux kernel \cite{ktau06}.
While this approach gave convincing results~\cite{nataraj2007ghost},
its lack of portability is a major drawback: the patches need to be
adapted to each version of the kernel, and it
required recompiling the kernel with them.

In this short paper, we
present a more portable, less invasive approach, using modern
technologies provided by the Linux kernel. 

Section \ref{sec:profiling} presents briefly the usual approaches
followed to obtain and collect performance information in
parallel applications.
Section \ref{sec:stap}
describes the approach we are presenting in this paper.
Section \ref{sec:eval}
provides a practical evaluation of our approach,
and section \ref{sec:conclu} concludes the paper.

\section{Traditional performance analysis approaches}
\label{sec:profiling}

Performance analysis can be done using two kinds of
measurement:
\emph{profiling} provides information on how much time
(and other metrics) is spent in parts of the program,
while \emph{tracing}
captures a timestamped series of
these events.
Both kinds of performance data are
collected by currently available tools
during the execution of the program,
for instance,
we can cite \textit{TAU}~\cite{Shende2006},
\textit{APEX}~\cite{huck2022broad},
HPCToolkit~\cite{Adhianto2010},
\textit{Scalasca}~\cite{geimer2010scalasca},
and \textit{Score-P}~\cite{mey2012score}.

Most HPC applications use programming interfaces that follow
a certain standard
for supporting performance tools.
Some of them define a \emph{profiling interface} where tools
can
connect.
For instance, the MPI interface defines a
\emph{name-shifted} version of
all the MPI routines.
These PMPI routines do exactly what the corresponding MPI routines
are documented to do,
in that,
the former
is called by the latter.
In this way,
performance analysis libraries can insert themselves
before and after
the public interface (the MPI routines)
\cite{vetter2005mpip}\cite{ramesh2017mpi}.
Other interfaces have been proposed to succeed to the PMPI interface,
such as QMPI~\cite{elis2019qmpi} and MPIT~\cite{clauss2011performance}. 

Some other programming environments define a \emph{callback interface}.
The idea is that
at specific points of the execution, a callback function is triggered.
These points can be when a task is created, when a parallel region
begins and ends,
and so on.
Performance profiling libraries implement these callback functions
and are loaded by the programming environment to replace
the default ones
(typically NULL functions).
We can cite OpenMP's OMPT interface~\cite{eichenberger2013ompt} and
its support in APEX \cite{huck2022broad}, and OpenACC's implementation
in Clacc and support in TAU~\cite{coti2020openacc}. Programming model
abstractions like Kokkos\cite{hammond2018profiling} and Raja\cite{beckingsale2019performance} have also included
callback APIs.

These approaches are used to measure the time spent in run-time
environments and their communication and execution routines.
It is also possible to measure the time spent in arbitrary portions
of the program.
In this case,
the code can be instrumented to insert
\emph{probes},
which can be done manually or using an automatic source instrumentation
tool such as TAU and PDT~\cite{lindlan2000tool} or a compiler-based
tool~\cite{coti2020openacc}.
Other tools instrument the binary itself~\cite{bernat2011anywhere} \cite{weidendorfer2008sequential}.
PerfStubs\cite{boehme2019case} provides a general purpose 
instrumentation API usage in C, C++ or Fortran
library or application code.

\section{Probing and instrumenting system calls}
\label{sec:stap}

System calls are the interface between user-level application and
the operating system: they are routines that can be called by the
applications to request operations that require higher privileges.
Hence, measuring the time spent in system calls measures the time
spent by the application requesting operations from the kernel. 
The techniques described above can be used to measure performance
at the application level outside of the system call, but they
can not see inside.

Our approach described below will make that possible.
It can be extended to other internal routines of the kernel,
such as the scheduler's internal routines.

\subsection{SystemTap}

\textit{SystemTap} is an infrastructure designed to monitor the Linux
Kernel's activity~\cite{eigler2006problem}.
It provides an interface (the tapset library) and a scripting
language to implement actions taken upon some specific events
in the kernel. Scripts are compiled by the stap compiler,
and loaded as modules of the kernel. 

SystemTap scripts probe kernel-space events that can be triggered
by instructions at a particular location in the kernel
(called \emph{synchronous events}), such as when a given function
is entered or at a tracepoint.
The event can also be
associated with constructs such as counters or timers,
not tied to a particular instruction or location in code.
These are
called \emph{asynchronous events}.
SystemTap does not require the user to have super-user privileges,
as long as it is included in the appropriate user group.

\textit{Ptrace} \cite{keniston2007ptrace} is a tool designed
specifically to measure the time spent by an application
(the \emph{tracee}) in system calls.
However, as mentioned by the authors of \cite{keniston2007ptrace},
it has some limitations and only provide some information, and
they explicitly recommend using one of the available interfaces
to implement a specific kernel-tracing tool.

\subsection{Design of STaKTAU}

Like many performance analysis tools, \textit{STaKTAU} is made
of two components:
\begin{itemize}
    \item a performance data collection tool;
    \item a user interface that gathers and displays the information
          collected in this data. 
\end{itemize}

Data is collected using SystemTap.
Every time a system call is entered, STaKTAU starts a timer;
when this system call is exited, STaKTAU ends this timer and stores
the data.

The data collected by our tool is in the kernel space and has to be
transferred to the user space to be
merged with other
user-level performance data.
The major challenge is to
do this efficiently and fast.
This constraint eliminates any solution that would write timing
information as soon as it is collected. 

One solution would be to store all the performance data during
the execution of the application and output it at the end.
However, this part is executed in kernel space, where
usable memory is limited.
As a consequence, we chose an intermediate solution:
the profiling data is stored in a buffer, and this buffer
is dumped once it is full and at the end of the execution.

K-TAU \cite{nataraj2007ghost} uses a segment of shared memory
to store data that can be collected by a user-space program.
By default, SystemTap can print information on the standard output,
or redirect it into a file.
This solution
takes advantage of performance optimization provided by the OS,
namely buffering.
Hence, we chose to dump the information on the standard output
of the script, and redirect this standard output info a file. 

This file can be later read by a
performance analysis tool.
An excerpt of a the output obtained profiling a small OpenMP program
is given below (the information concerning 2 threads only is copied
here). The total execution time is the time between the first system call and the call to {\tt exit}.

\begin{Verbatim}[fontsize=\small]
 ----- TID 40705 -----
call                    | time
---------------------------------
STaKTAU application    | 1680935665981760030
rt_sigsuspend           | 246097
rt_sigaction            | 5492
rt_sigprocmask          | 223200
alarm                   | 3282
execve                  | 417912
brk                     | 8862
arch_prctl              | 4727
mmap2                   | 256817
access                  | 9123
openat                  | 37108
fstatat                 | 24482
close                   | 8693
read                    | 7813
pread                   | 11211
set_tid_address         | 2383
set_robust_list         | 2165
mprotect                | 251824
prlimit64               | 3368
munmap                  | 11140
getrandom               | 3069
getdents                | 23968
sched_getaffinity       | 3931
futex                   | 12486935
write                   | 505633
----- TID 40739 -----
call                    | time
---------------------------------
STaKTAU application    | 1680935665984544508
set_robust_list         | 4760
rt_sigprocmask          | 4213
futex                   | 17912269
\end{Verbatim}

\section{Evaluation}
\label{sec:eval}

The practical evaluation of our tool involved two aspects:
is the approach doable at all, and how much overhead does it have
on the execution.

We conducted this performance evaluation on a machine from the
Grisou cluster of Grid'5000 \cite{grid5000}.
The nodes in this cluster feature two Intel Xeon E5-2630 v3	CPUs,
with 8 hyperthreaded cores each (hence 32 hardware threads per node),
and 128 GB or memory.
We deployed an Ubuntu 22.04 environment with a Linux 5.10.0-38 kernel,
gcc 11.3.0, and we compiled SystemTap from their Git Repository
(hash {\tt 6f9a9bc6daaa9ea44fea7067c45f218397dc3d96},
(version 4.9/0.186, release-4.8-55-g6f9a9bc6-dirty).

When events happen faster than SystemTap can output them,
it
drops
them and displays a warning giving the number of missed events.
Our hypothesis was that if we output information after every
system call,
too many events will be queued and SystemTap will start missing some.

We noticed this happening very quickly, on simple parallel
applications using OpenMP.
The main question was then: can we store the information in
a buffer large enough to avoid missing events, but small enough
to be allowed in the kernel?
Choosing a buffer size of 2048 elements
was enough to never miss any event on the 32-core machine we used.

The second point is the overhead.
We measured it using two benchmarks: the first one is a
naive computation of $\pi$ such as the one presented in many
OpenMP tutorials, using a reduction.
This application is doing little computation and, in fact, does
not scale at all on the machine we used.
It is mostly
exercising the OpenMP infrastructure
(i.e., spawning and synchronizing threads).

The other application we used is a ray tracing application.
It is spending most of its time computing the ray tracing
operations in two nested loops, the outer being parallelized
with OpenMP, with file operations at the beginning and the
end of the execution.
The goal of this benchmark is to show the overhead of STaKTAU
on a more computation-intensive application, with a use of the
operating system more typical from HPC applications.

Table \ref{tab:perf} presents the execution time of these two
applications using 32 threads, with and without kernel-level
profiling.
We ran each application 10 times, on the same machine.
As expected, the $\pi$ computation benchmark has a larger
overhead, since it is spending a larger portion of its execution
time in system calls.
On the ray tracing application, on the other hand, we notice
that the overhead is negligible (the difference is included in
the standard deviation). 

In order to have an idea of the relative logging intensity of
these two applications, we are also showing the size of the output
trace file.
We can see that the $\pi$ benchmark generated about 1~kB per
millisecond, whereas the ray tracing benchmark generated a bit less
than 3~B per millisecond.
This observation is consistent to the explanation of the higher
overhead with the $\pi$ benchmark.

\begin{table}[!ht]
    \centering
    \begin{tabular}{l|c|c|c}
    Benchmark & Execution time (s) & Std dev & Trace file size\\
    \hline
    Pi \\
  Vanilla:&	 0.037 &	 0.003 \\
  STaKTAU: & 	 0.062 &	 0.004 & 65\,236~B\\
  Overhead:&	 169.2\%\\
\hline
Ray tracing:\\
  Vanilla:&	 20.946 &	 0.113\\
  STaKTAU: & 	 20.823 	& 0.121\ & 82\,332~B\\
  Overhead:	 & 99.2\%\\
  \hline
    \end{tabular}
    \caption{Overhead of kernel-level performance profiling on two OpenMP benchmarks.\label{tab:perf}}
\end{table}

\section{Conclusion}
\label{sec:conclu}

In this short paper, we have presented a kernel-level profiling tool
for high performance applications.
It relies on SystemTap, a framework for kernel activity monitoring.
Major advantages of this approach is that it is portable across
versions of the Linux kernel, it does not require recompiling the
kernel, and it does not require root privileges.
Performance evaluation on OpenMP applications shows that our tool
as little overhead on the execution time, provided that the
application is actually computing and not spending most of its
execution time in system calls.

STaKTAU is a standalone tool: it gives information on the time spent by the application in the kernel. In order to have this information as well as other performance information relevant to the execution, it has to be interfaced with a performance analysis tool with a broader scope. Moreover, including it in a hierarchical performance analysis tool would give more precise information: in which parts of the code does the application spend time in the kernel? Is it in memory movements handled by an external library, or is it in the code written by the application developers? 

STaKTAU is available for download at \url{www.github.com/coti/STaKTAU}.

\printbibliography 

\end{document}